# Improving prediction of students' performance in Intelligent Tutoring Systems using attribute selection and ensembles of different multimodal data sources

Wilson Chango, Rebeca Cerezo, Miguel Sanchez-Santillan, Roger Azevedo, Cristóbal Romero

1 Pontifical Catholic University of Ecuador, Ecuador (wilson.chango@pucese.edu.ec)
2 University of Oviedo, Spain (sanchezsmiguel@uniovi.es, cerezorebeca@uniovi.es)
3 University of Central Florida, USA (roger.azevedo@ucf.edu)
4 University of Córdoba, Spain (cromero@uco.es)

## ABSTRACT

The aim of this study was to predict university students' learning performance using different sources of data from an Intelligent Tutoring System. We collected and preprocessed data from 40 students from different multimodal sources: learning strategies from system logs, emotions from face recording videos, interaction zones from eye tracking, and test performance from final knowledge evaluation. Our objective was to test whether the prediction could be improved by using attribute selection and classification ensembles. We carried out three experiments by applying six classification algorithms to numerical and discretized preprocessed multimodal data. The results show that the best predictions were produced using ensembles and selecting the best attributes approach with numerical data.

## KEYWORDS

Predicting academic performance, Intelligent Tutoring Systems, multisource data, multimodal learning, data fusion.

## 1 Introduction

The rapid growth of technology has meant that computer learning has increasingly integrated artificial intelligence techniques in order to develop more personalized educational systems. These systems are known as Intelligent Tutoring systems (ITS).

MetaTutorES (Cerezo, Esteban, et al., 2020; Cerezo, Fernández, et al., 2020), a Spanish adaptation of MetaTutor (Azevedo, 2009) is an ITS designed to detect, model, trace, and foster students' self-regulated learning while learning various science topics (e.g., by modeling and scaffolding metacognitive monitoring, facilitating the use of effective learning strategies, and setting and coordinating relevant learning goals). The system uses human-like avatar technology that allows pedagogical agents to track student behavior and provide interaction on this basis. Tracking students' behavior is also a powerful research tool used to collect data on students' cognitive, metacognitive, affective, and motivational processes deployed during learning (Azevedo et al., 2011; Greene & Azevedo, 2010; Harley et al., 2014). These different data sources can be fused and mined to to reveal learning-related information such as student performance. In this regard, Educational data mining (EDM) and Learning Analytics (LA) can be applied to understand educational processes using information extracted from educational data, which is then used to improve the educational process and the quality of learning (Cristobal Romero & Ventura, 2020).

One of the oldest and most commonly studied issues in EDM/LA is the prediction of learners' performance. It is still a challenge to predict student learning achievement in ITSs using Multimodal Learning Analytics (MLA) with learning data from different sources and doing a single analysis (Blikstein & Worsley, 2016). MLA uses log-files and gaze data, biosensors, interactions with videos, audio and digital documents, and any other relevant data source to measure and understand the learning process.

One important issue in MLA is how to combine, or fuse, the data extracted from various sources/modalities in order to provide a better, more comprehensive view of teaching-learning processes (Bogarín et al., 2018; Chango et al., 2021). The most common and simplest data fusion approach for combining all the data sources is to build a machine-learning classifier from the summary statistics produced from each of the data sources. An important task when fusing data is to reduce the dimensions of the variables/attributes and to identify the most fruitful feature sets. Feature selection algorithms are normally used in data fusion for classification problems in order to reduce the data dimensions and produce the best results (Jesus et al., 2017). Finally, classification ensembles have demonstrated very good results in predicting student academic performance from multimodal data sources (Adejo & Connolly, 2018).

In this paper we perform a classification task, predicting the value of a categorical/nominal attribute (the class or final knowledge status of the student (Pass, Fail) based on other attributes (the predictive attributes from various available data sources). We propose applying classification algorithms, feature selection algorithms, and ensembles to data gathered from a variety of sources (learning strategies from ITS logs, emotions from face recording videos, and interaction zones from eye tracking) in order to predict the students' final performance in the ITS. In this sense, the ultimate contribution of this study is to analyze the learning process through resources, allowing a more personalized response to each learner.

The research questions posed by this study are:

Question 1.- Can attribute selection and classification ensemble algorithms improve the prediction of students' final performance from our ITS data?

Question 2.- How useful are the models produced and what are the best variables to help teachers understand how to predict students' final performance in the ITS?

This paper is organized as follows. The first section covers the background of the related research area of MLA. Subsequently, we describe the proposed methodology, the data used, and how it was preprocessed. Then, we describe the experiments we performed and the results they produced. Finally, we discuss the implications, conclusions, and lines for future research.

## 2   Background

MLA aims to combine different sources of learning traces into a single analysis, it is a subfield of EDM related to multi-view and multi-relational data and data fusion. It aims to understand and optimize learning in digital where the use of videos is currently consolidated, from traditional courses to mixed and online courses (Chan et al., 2020). MLA can generate distinctive insights into what happens when students create unique solution paths to problems, interact with peers, and act in both physical and digital environments. It has become increasingly broadly applied in both digital and in real-world scenarios where interactions are not solely mediated through computers or digital devices (Blikstein & Worsley, 2016). In MLA, learning traces are extracted not only from log-files but also from digital documents, recorded video and audio, pen strokes, position tracking devices, biosensors, and any other data sources that could be useful for understanding or measuring the learning process. Below, we describe the data sources used in the present study.

**Learning strategies from ITS logs**

There is empirical evidence about performance prediction through computer learning environment log data (Cerezo et al., 2016; Lerche & Kiel, 2018; Li & Tsai, 2017), including predicting performance in offline courses from logs of online behavior (Zhong et al., 2015). As computer-based learning environments, ITSs allow us to see what learning strategies users deploy while they are studying, and are part of a new trend in the measurement of learning in general, and self-regulated learning in particular—the so called third wave—, characterized by combined use of measurement and Advanced Learning Technologies (Panadero et al., 2016). These performance analytics include data on the student's performance and different learning metrics. Example include completion time, successful or unsuccessful completion of assignments, speed of task resolution, the number of attempts or failures, and the complexity of the problem-solving process (Crescenzi-Lanna, 2020). All of these data are normally produced by the computer during the student's interaction with the learning environment and are stored in database or log files (Cristóbal Romero et al., 2008). This technology overcomes the limitations of self-report methodology, making it possible to detect, model, trace, and encourage students' learning, with the added benefit of not interfering with student activity, because even though a huge amount of data is generated, it is processed automatically by the computer.

**Interaction zones from eye tracking**

Eye-tracking devices provide information that can be used to infer the student's attention level, engagement, preference, or understanding. It provides an understanding of what attracts immediate attention, which target elements are ignored, what order elements are noticed in, and how elements compare to others (Cerezo, Fernández, et al., 2020). Gaze data can provide very useful, accurate information for predicting student learning during interaction with ITSs (Bondareva et al., 2013), and multiple researchers have suggested that the duration of fixations are indicators of cognitive processing during learning (Antonietti et al., 2015).

There are different options for collecting eye-tracking data such as saccade amplitude and direction change, and fixations, etc. (Crescenzi-Lanna, 2020). In the current study, we are interested in analyzing fixations, particularly the number of fixations in areas that could be related to the learner's final performance. For that purpose, we defined three Areas of interest (AOIs) in our ITS interface: AOI1 Learning session timer, AOI2 ITS agent/avatar, and AOI3 Supporting image/graphics content. These are areas of interest because, in terms of the interface configurations, fixations on AOI1 may denote time management or resource management strategies, while reduced or excessive fixations on AOI1 might indicate poor time management skills. Fixation on AOI2, the agent, would show that the participant is making use of the prompts and feedback provided by the agents during the learning session and has established an interaction with the agent. Fixations on AOI3 may point to participants using a strategy of coordinating information sources (text-images), associated with learning gains (Azevedo, 2009; Cerezo, Fernández, et al., 2020).

**Emotions from face recording videos**

Emotions are a critical component of learning and problem solving, especially when it comes to interacting with computer-based learning environments (Harley et al., 2015), and there is a relationship between negative learning emotion and learning performance (Chen & Wang, 2011). In this context, studies from affective computing literature suggest that facial expressions may be the best single method for accurately identifying emotional states (D'Mello & Kory, 2012). Techniques for automatic detection of emotions (Blanchard et al., 2009) are capable of isolating a learner's mood via artificial intelligence facial recognition systems, and there are tools available that can process video data, such as the Microsoft Emotion API (2019), Face API (2019), and Affectiva (2019).

As far as we are aware, no previous studies have examined whether the emotion recognition output of these tools is powerful enough to be used to predict student performance in ITS. However, including the learner's emotional states may help enhance ITS quality and efficacy. Previous research has indicated that academic emotions are significantly related to students' motivation, learning strategies, cognitive resources, self-regulation, and academic achievement (Pekrun et al., 2011).

In previous studies, student emotions as recognized by an API during a learning session with an ITS have been used as the sole data source for predicting the student's final performance. The best models demonstrated a prediction accuracy of 63.82% and 0.67 AUC, figures that we aim to improve on by using more student features and variables from various multimodal data sources, together with ensembles and selection of the best attributes.

## 3  Proposal

The current study proposes a two-stage methodology for predicting students' final performance from multimodal data (see Figure 1).

**Figure 1. Proposed methodology for predicting students' performance from multiple data sources.**

As Figure 1 shows, the two main stages in our methodology are:

- First stage. Collecting data from various sources: learning strategies from Metatutor logs, number of fixations from gaze data, and emotions from face recording videos. It also includes some pre-processing tasks (anonymization, attribute normalization and discretization, and format transformation) to generate numerical and categorical datasets.

- Second stage. Using different data fusion approaches: merge all attributes; selection of the best attributes, and ensembles of several white box classification algorithms. Finally, the predictions produced by the models are compared in order to find the best model and attributes to be used to predict the students' final performance.

## 4   Data

Data were collected from 40 undergraduates (mean age = 23.58; SD = 8.18; 17 men and 23 women) enrolled at a public university in the north of Spain. The undergraduates participated in the study voluntarily and learned about a complex science topic (the circulatory system) while interacting with the MetaTutorES ITS (Cerezo, Esteban, et al., 2020; Cerezo, Fernández, et al., 2020), a computerized learning environment. The students in the sample were studying in a variety of different knowledge areas: education, psychology, economics, law, philosophy, nursing, telecommunication, electrical engineering, geomatics, physics, and civil navy. Most students in the sample were first-year undergraduates, but there were also second-years, third-years and masters.

### Gathering Data

We gathered information from four ITS data sources: learning strategies from MetatutorES logs, emotions from face videos, fixation from eye tracking, and performance from the content knowledge test. The data collected was produced spontaneously from interactions with the MetaTutorES ITS during a session lasting from two-and-a-half to three hours. The data collection for the study was developed and managed in line with the ethical research principles of the Declaration of Helsinki and the protocol was approved by the research ethics committee of the Principality of Asturias and the University of Oviedo.

#### 4.1.1   Learning strategies from Metatutor logs

Throughout each learning session, learner interaction with the ITS was logged in a log file unique to each learner. The learning environment is made up of information in text, charts, and images, through which students learn about the circulatory system. The system logs each user action and interaction with the learning environment and the study. Each line of a log represents an event or participant action in the learning environment and contains the timestamp of the event, the triggered event, the identifier of the theoretical content that the learner is studying and optional information related to that event.

For the present study, three variables were extracted from the log files: SummAll: The number of times that the learner wrote a *Summary* about the content they were studying, discarding the events in which they did not add any new information, e. g. After spending time reading the page about the role of the heart in the circulatory system, the user summarizes the reading; COIStotalFreq: The number of times the learner enlarged the image associated with the content being studied for at least fifteen seconds *Coordinating Information Sources* (e.g. drawing and text), e.g. Spend time studying about the heart and open the associated image. PKAtotalFreq: *Prior Knowledge Activation* is the number of times that the learner, after navigating to previously unvisited content, writes their prior knowledge about the new content. A correlate for when the student searches in their memory for relevant prior knowledge either before beginning task performance or during task performance., e.g. The student opens a page and, before reading, writes everything they already know about the topic on that page.

### 4.1.2  Emotions from Face Recording Video

During the learning session a video of the participants face was recorded using a web cam which was subsequently analyzed using a desktop app. Each participant's full session was recorded, the webcam on the computer was adjusted to the participant's position at the beginning and they were asked to sit facing forward and be as neutral as possible, although their facial expressions were expected to vary during the session. We asked participants to tie their hair back, make sure there was nothing around their neck, remove their glasses, and remove chewing gum if necessary to have the best conditions for the recording. The learning session videos were analyzed using Microsoft Emotion API (2019 automatic facial recognition software). The API classifies facial expression in eight emotion classes: anger, contempt, disgust, fear, happiness, neutral, sadness, and surprise. These emotions are understood to be cross-culturally and universally communicated with specific facial expressions (Arora et al., 2018).  We developed our specific application to use Microsoft Emotion API in local mode (see Figure ). Participants tended to experience all of emotions the system detects during the session, but we were able to produce a general index for each participant giving information about the general pattern. The analysis gave us at least one predominant emotion during the learning session from frame of student video, and there were a large number of frames (1 frame per second) for each student in every session. The confidence (values between 0 and 1) gives the likelihood for each class of emotion.

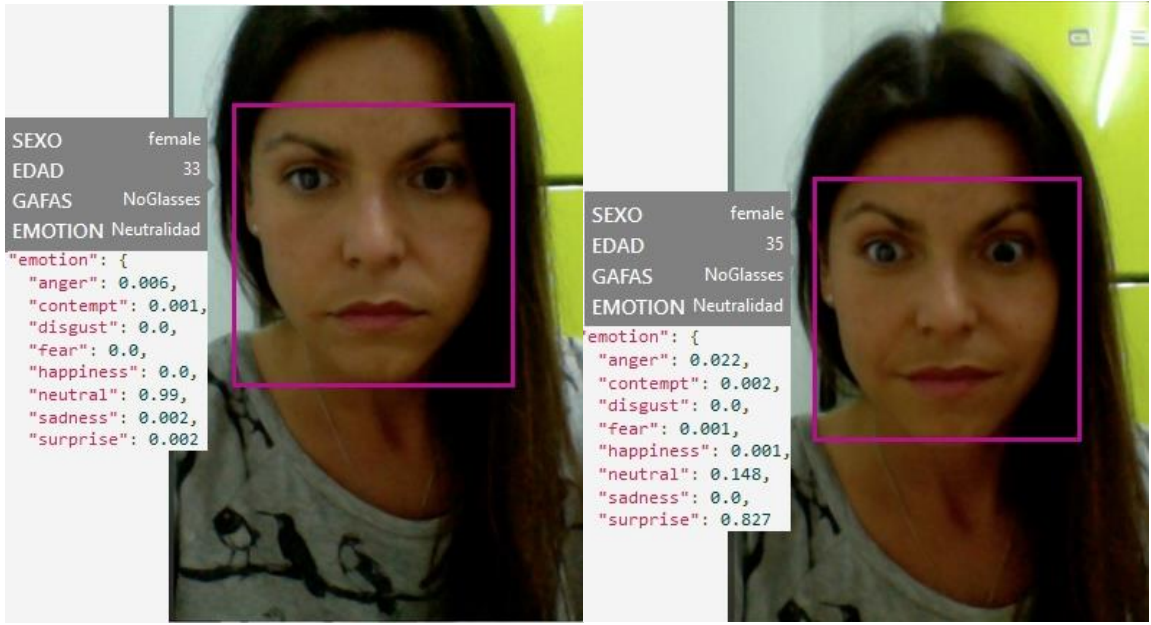

**Figure 2. Examples of image of emotion recognition (the left-hand column shows the emotion trend).**

### 4.1.3 Interaction Zones from Eye tracking

Data from each learner was collected throughout the session using the screen-based eye tracker RED500 (https://imotions.com/hardware/smi-red500/). We used SMI's BeGaze software in order to process the fixations on the learning environment AOIs. BeGaze performs the calculation automatically, identifying a fixation if a learner stares at an AOI for at least 80ms with a maximum dispersion of 100px.

For the present study, we extracted three variables related to learner fixation on three AOIs (See Figure 3). AOI1 The learning session timer (number of times the learner focused their attention on the area showing the time left in the learning session), which may denote time management or resource management strategies, while reduced or excessive fixations on AOI1 might indicate poor time management skills . AOI2 ITS agent/avatar (number of times the learner focused their attention on the area where the pedagogical agents appear). This variable may show that the participant is taking advantage of the prompts and feedback provided by the agents during the interaction in response to participants' goals, behaviors, self-evaluations, and progress. However, it must be considered carefully, because learners may not always need to look at an agent to process their audio prompts and feedback (Bondareva, et al. 2013). AOI3 Images/graphics supporting content (number of times the learner focused their attention on the area covered by the images related to the learning session contents). This variable may indicate integration contributing to information processing (Mason et al., 2013).

**Figure 3. Map of AOIs in the ITS.**

### 4.1.4  Final mark from Test/Quiz

During the session and at the end of the session, each subject was tested about the learning content, giving a final performance value between 0 and 10, with 10 being the highest performance. There was a pretest about prior knowledge of the content at the beginning of the session, and the final performance was corrected based on that.

**Preprocessing Data**

We preprocessed all of the data in the aforementioned Excel files (Cristóbal Romero et al., 2014). Firstly, the data were anonymized, then the input attributes were normalized/rescaled, the output attributes and input attributes were discretized, and finally the format was transformed.

### 4.1.5  Anonymizing

Student anonymity and privacy was maintained but the information in the four Excel files was linked to the same subject using anonymized coding. We implemented a basic solution, using a randomly generated number as a user ID rather than the users' names, and replaced the students' names with the ID in the four Excel files.

### 4.1.6  Normalizing

We adjusted all of the input values, which used different scales, to a single common scale. This was necessary because the original values had a variety of ranges. Normalization is a data transformation where the attribute values are scaled so as to fall within a specified range, such as -1.0 to 1.0, or 0.0 to 1.0. Normalization helps to prevent attributes with large ranges from outweighing attributes with smaller ranges. In this case we rescaled/normalized all of the input attribute values to the same range [0-1] by using the well-known Min-Max method, which is a linear transformation of the original data using the formula: $Z_i = X_i - min(X) / max(X) - min(X)$, where X=(x$_1$,...,x$_n$) and $Z_i$ is now the $ith$ normalized data.

### 4.1.7  Discretizing

Discretization divides numerical data into categorical classes that are more user-friendly than precise magnitudes and ranges. It reduces the number of possible values of the continuous feature and provides a view of the data that is easier to understand. Generally, discretization smooths out the effect of noise and enables simpler models, which are less prone to overfitting. We discretized all the input attributes in order to have the same variables in both numerical and categorical formats. To do that, we used equal-width binning with the following 3 bins: *LOW*, *MEDIUM* and *HIGH*. Equal-width binning divides the range of possible values into *N* sub-ranges of the same size in which: *bin_width = (max value − min value) / N*.

We also discretized the output attribute or class to predict (the students' final academic performance or status).  We used a manual discretization with the user directly specifying cut-off points. In our case, the class had the following 2 values and cut-off points:

- PASS: Students who scored 5 out of 10 or better in the final performance test. In our case, this was 21 out of 40 students (52.50%).
- FAIL: Students who scored less than 5 out of 10 in the final performance test. In our case, this was 19 out of 40 students (47.50%).

### 4.1.8  Transforming

Finally, we converted the files from Excel to CSV (Comma-separated values) files. CSV is a delimited text file that uses a comma to separate values. Each line of the file is a data record. Each record consists of one or more fields, separated by commas. We transformed each of the two versions of the four Excel files (numerical and categorical values) into two CSV files because they can be directly opened and used by the WEKA data mining framework that we used in the experiments. We used the WEKA (Waikato Environment for Knowledge Analysis) data mining framework (Witten et al., 2011) to predict student performance. WEKA provides a collection of algorithms for data analysis and predictive modeling, together with graphical user interfaces for easy access to these functions.

### 5  Experiments

We carried out three different experiments using three different approaches and six classification algorithms with the preprocessed numerical and discretized data to predict student performance in the ITS (See Figure 4).

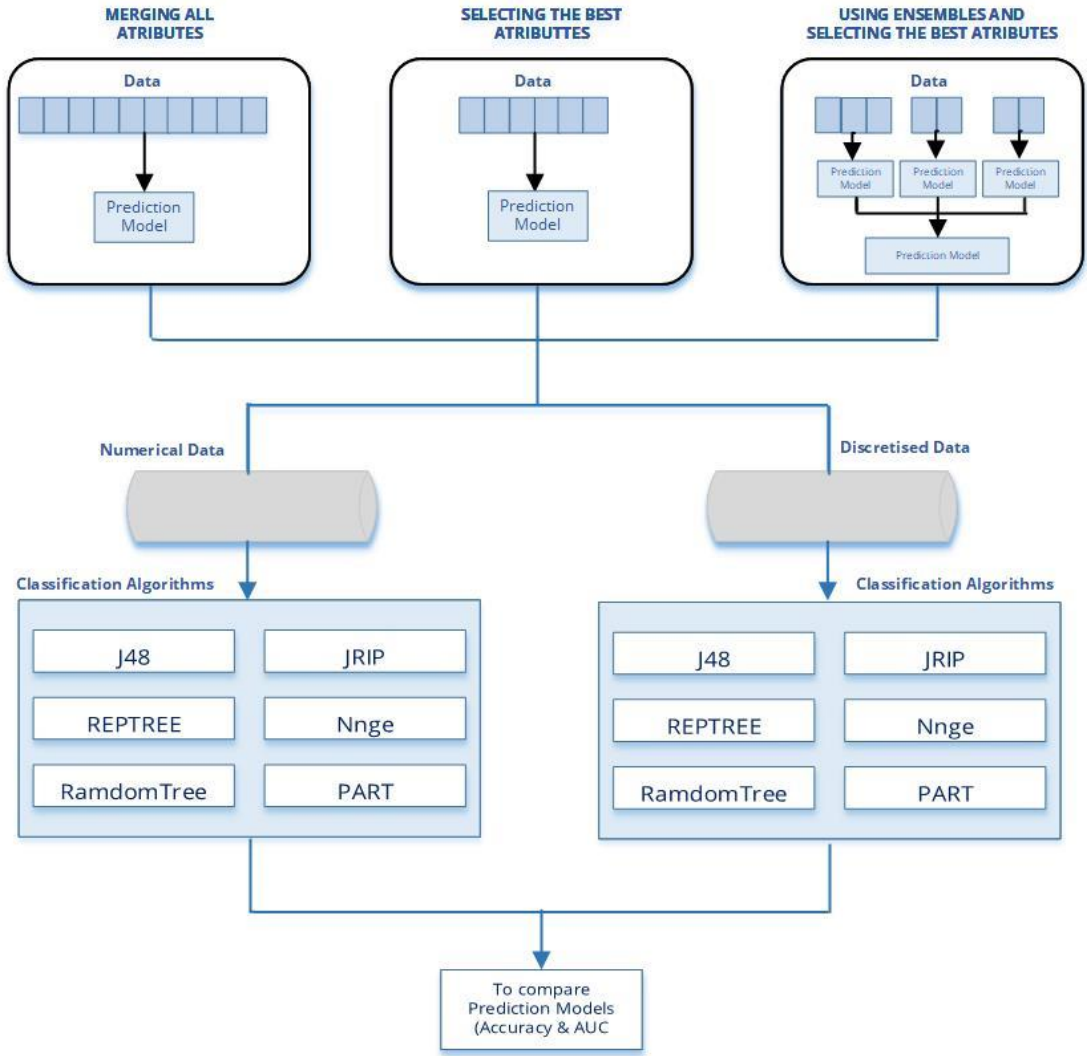

**Figure 4. Visual description of the experiments**

We used two types of white box classification models: Rule induction algorithms and decision trees. The models produced by these algorithms (IF-THEN rules from decision trees) are simple and clear, and so are easy for humans to understand. IF-THEN classification rules provide a high-level knowledge representation that is used for decision making, while decision trees can also be converted into a set of IF-THEN classification rules. In our experiments, we selected six well-known classification algorithms integrated in the WEKA data mining tool (Witten et al., 2011): three decision tree algorithms (J48, REPTree and RandomTree) and three rule induction algorithms (JRip, Nnge and PART). We executed these algorithms using a k-fold cross-validation (k=10) and Accuracy and Area under the ROC curve as evaluation metrics for classification:

- **Accuracy (ACC)** is the most commonly-used traditional method for evaluating classification algorithms. It provides a single-number summary of performance. In our case, it is obtained by the equation: Acc=$\frac{\text{Number of students correctly classified}}{\text{Total number of students}}$. This metric shows the percentage of correctly classified students.

- **Area under the ROC curve (AUC)** measures the two-dimensional area underneath the entire Relative Operating Characteristic (ROC) curve. The ROC curve allows us to find possibly optimal models and discard suboptimal ones. AUC is often used when the goal of classification is to obtain a ranking because ROC curve construction requires a ranking to be produced.

## Experiment 1: Merging all attributes

In experiment 1 we applied the classification algorithms to a single file with all the attributes of the three different data sources merged. We created two different numerical and discrete/categorical CSV files. Each dataset had fifteen input attributes (in numerical or discrete format) and only one output attribute or class. Finally, we executed six classification algorithms on the two summary datasets, producing the results (%Accuracy and ROC Area) shown in Table 1.

**Table 1. Results produced by merging all attributes**

|  | NUMERICAL DATA | | DISCRETIZED DATA | |
|---|---|---|---|---|
|  | **% Accuracy** | **AUC** | %Accuracy | AUC |
| Jrip | 72.50 | 0.69 | 72.50 | 0.65 |
| Nnge | 62.50 | 0.61 | 62.50 | 0.62 |
| PART | **80.00** | 0.79 | 67.50 | 0.69 |
| **J48** | **80.00** | **0.80** | 70.00 | 0.67 |
| REPTree | 72.50 | 0.74 | 67.50 | 0.61 |
| Randomtree | 70.00 | 0.70 | 72.50 | 0.69 |
| Avg. | 73.33 | 0.72 | 68.75 | 0.66 |

Table 1 shows that the best results (highest values) were produced by Part (80.0 %Acc) and J48 (80.00 %Acc and 0.80 AUC) algorithms with numerical data. In fact, on average, most of the algorithms exhibited slightly improved performance in both measures when using numerical data.

## Experiment 2: Selecting the best attributes

In Experiment 2, we applied the classification algorithms to a single file with only the best attributes. Firstly, we applied attribute selection algorithms to the summary files from the Experiment 1 in order to eliminate redundant or irrelevant attributes. We used the well-known *CfsSubsetEval* (Correlation-based Featured Selection) method provided by the WEKA tool. This method selects the features that are more strongly correlated with the class. Starting from our initial 15 input attributes, we produced two sets of 2 optimal attributes for the numerical datasets and 5 optimal attributes (see Table 2) for the discretized datasets.

**Table 2. Results of the attribute selection with CLASSIFIERSUBSETEVAL.**

| Dataset | # selected features | Name of Selected features |
|---------|--------------------|--------------------------|
| Numerical | 2 | Metatutor.SummAll |
| | | Metatutor.COIStotalFreq |
| Discretized | 5 | Metatutor.SummAll |
| | | Interaction.AOI1FixCount |
| | | Interaction.AOI3FixCount |
| | | Emotion.anger |
| | | Emotion.happiness |

Following that, we executed the six classification algorithms with the two new summary datasets, producing the results (%Accuracy and ROC Area) shown in Table 3**¡Error! No se encuentra el origen de la referencia.**.

**Table 3. Results obtained when selecting the best attributes.**

| | NUMERICAL DATA | | DISCRETIZED DATA | |
|---|---|---|---|---|
| | % Accuracy | AUC | %Accuracy | AUC |
| Jrip | 77.50 | 0.81 | 77.50 | 0.68 |
| Nnge | 80.00 | 0.80 | 75.00 | 0.75 |
| PART | 77.50 | 0.77 | 70.00 | 0.67 |
| J48 | 77.50 | 0.80 | 77.50 | 0.76 |
| REPTree | 80.00 | 0.78 | 70.00 | 0.63 |
| **Randomtree** | **82.50** | **0.82** | 75.00 | 0.77 |

Table 3 shows that the best results (highest values) were produced by Randomtree (82.50 %Acc and 0.82 AUC) algorithms. Again, on average most of the algorithms exhibited slightly improved performance in both measures when using numerical data.

**Experiment 3: Using ensembles and selecting the best attributes**

In Experiment 3 we applied an ensemble of classification algorithms to the best attributes from each different data source. Firstly, we selected the best attributes for each of the three different datasets, again using the well-known *CfsSubsetEval* attribute selection algorithm. This gave the list of attributes shown in Table 4.

**Table 4. Results of attribute selection with CFSSubsetEval.**

| Dataset | Type | # selected features | Name of Selected features |
|---------|------|---------------------|---------------------------|
| Metatutor | Numerical | 1 | Metatutor.SummAll |
| | Discretized | 1 | Metatutor.SummAll |
| Interaction | Numerical | 1 | Interaction.AOI6FixCount |
| | Discretized | 2 | Interaction.AOI6FixCount Interaction.AOI1FixCount |
| Emotion | Numerical | 1 | Emotion.surprise |
| | Discretized | 1 | Emotion.fear |

Following that, we applied an ensemble or combination of multiple classification base models by using the well-known Vote (Kuncheva, 2014) for automatic combining several machine learning algorithms provided by WEKA. Vote combines the probability distributions of these base learners. It produces better results than individual classification models, if the set classifiers are accurate and diverse. It has demonstrated better results than homogeneous models for standard datasets.
We executed the six classification algorithms as base or individual classification models of our Vote method with the previously described numerical and discretized datasets. Table 5 shows the results (%Accuracy and ROC Area).

**Table 5. Results from using ensembles and selecting the best attributes.**

| | NUMERICAL DATA | | DISCRETIZED DATA | |
|---|---|---|---|---|
| | % Accuracy | AUC | %Accuracy | AUC |
| Jrip | 82.50 | 0.88 | 82.50 | 0.86 |
| Nnge | 80.00 | 0.87 | 65.00 | 0.66 |
| PART | 80.00 | 0.84 | 75.00 | 0.78 |
| J48 | 82.50 | 0.86 | 80.00 | 0.84 |
| **REPTree** | **87.50** | **0.88** | 80.00 | 0.82 |
| Randomtree | 82.50 | 0.88 | 75.00 | 0.74 |

Table 5 shows that the best results (highest values) were produced by REPTree (87.50 %Acc and 0.88 AUC). On average, most of the algorithms again exhibited slightly improved performance in both measures when using numerical data.

## 6. Discussion

Below, we address the two initial research questions by discussing the results from our four experiments.

*Question 1*

*Can attribute selection and classification ensemble algorithms improve the prediction results of student final performance from our ITS data?*

We used three different data fusion approaches and six white-box classification algorithms to answer this question. Table 6 shows that the average prediction performance (Average of % Accuracy and AUC) of the classification algorithms increased with each new approach.

**Table 6. Average results from the three data fusion approaches.**

| Average | **NUMERICAL DATA** | | DISCRETIZED DATA | |
|---|---|---|---|---|
| | **% Accuracy** | **AUC** | %Accuracy | AUC |
| Merging all attributes | 73.33 | 0.72 | 68.75 | 0.66 |
| Selecting the best attributes | 79.16 | 0.80 | 74.16 | 0.71 |
| **Using ensembles and selection of the best attributes** | **82.50** | **0.87** | 76.25 | 0.78 |

We first applied a traditional approach for merging all the attributes from the different data sources directly. This initial approach gave reasonable results (accuracy higher than 70% and AUC higher that 0.7) from numerical data. Our second approach selected the best attributes for each dataset. This was an improvement on the first approach (79% accuracy and 0.8 AUC). Finally, the third approach improved on the second approach and gave the best result by using ensembles and selection of the best attributes (82% accuracy and 0.87 AUC). In all the approaches the average values were higher when using numerical than discretized data.

However, we were unable to find a single best algorithm that would win in all cases in our experiments. This is logical and in line with the No-Free-Lunch theorem (Wolpert, 2002), in which it is generally accepted that no single supervised learning algorithm can beat another algorithm over all possible learning problems or different datasets. In the first experiment, the algorithm that produced the highest prediction values was J48 (80.00 %Acc and 0.80 AUC), in the second experiment it was Randomtree (82.50 %Acc and 0.82 AUC), and REPTREE produced the highest prediction values of %Acc (87.50) and AUC (0.88) when using an ensemble and selection of the best attributes from the discretized data in the fourth experiment.

*Question 2*

> *How useful are the models produced and what are the best variables to help teachers understand how to predict students' final performance in the ITS?*

To answer this question, we will demonstrate and describe the meaning of the prediction model that produced the highest values of Accuracy and AUC in each of our 3 experiments.

In experiment 1, the prediction model producing the best prediction was produced by the J48 algorithm using discretized data (see Table 7)

**Table 7. J48 decision tree produced when merging all attributes.**

| |
|---|
| If Metatutor.SummAll > 0.25 Then PASS |
| If Metatutor.SummAll <=0.25 AND Emotions.surprise <=0.061227 Then FAIL |
| If Emotions.surprise > 0.06 AND Interaction.AOI3FixCount<=0.04 Then PASS |
| Else FAIL |
| Number of Rules:           4 |

This prediction model (see Table 7) has 4 rules. The first rule shows that the students who have scores higher than 0.25 in SummAll in Metatutor PASS the course. The second rule shows that if students have a score lower than 0.25 in SummAll in Metatutor and a surprise emotion lower than 0.06, then they FAIL the course. The third rule shows that if students have a surprise emotion higher than 0.06 and a value of AOI2FixCount lower than 0.04 in the Agent/avatar zone, then they PASS the course. Finally, the remaining students are classified as FAIL.

In experiment 2, the prediction model that produced the highest prediction values used the Randomtree algorithm with numerical data (see Table 8).

**Table 8. Randomtree pruned tree produced when selecting the best attributes.**

| |
|---|
| If Metatutor.SummAll < 0.28 |
| |   Metatutor.COIStotalFreq < 0.04 Then Pass |
| |   IF Metatutor.COIStotalFreq >= 0.04 |

| | IF Metatutor.SummAll < 0.03

| | | Metatutor.COIStotalFreq < 0.56 Then Fail

| | | IF COIStotalFreq >= 0.56

| | | | IF COIStotalFreq < 0.66

| | | | | Metatutor.COIStotalFreq < 0.59 Then Pass

| | | | | Metatutor.COIStotalFreq >= 0.59 Then Pass

| | | | Else Metatutor.COIStotalFreq >= 0.66 Then Fail

| | Else IF Metatutor.SummAll >= 0.03

| | | Metatutor.SummAll < 0.16 Then Pass

| | | Metatutor.SummAll >= 0.16 Then Fail

Else Metatutor.SummAll >= 0.28: Pass

Size of the tree : 15

This prediction model (see Table 8) consists of 7 IF-THEN rules. In all these rules, the two most frequent attributes are the summary strategies (SummAll) and the frequency of use of the user coordination of information sources strategy (COIStotalFreq). It is also important to note that in this model the predictions of students passing or failing was not influenced by any emotions or interaction zones.

In experiment 3, the prediction model that produced the highest prediction values used the RepTree algorithm with numerical data (see Table 9).

**Table 9. RepTree decision trees produced using ensembles with selecting the best attributes.**

REPTree (Metatutor)

============

If Metatutor SummAll >= 0.03 Then Pass

Else Fail

Size of the tree: 3

REPTree (Interaction)

============

If Interaction.AOI3FixCount >= 0.29 Then Pass

Else Fail

Size of the tree: 3

REPTree (Emotion)

============

If Emotion.surprise >= 0.05 Then Pass

| Else Fail |
|---|
| Size of the tree : 3 |

This prediction model (see **¡Error! No se encuentra el origen de la referencia.**) is a combination of three different models showing that the behavior of students in relation to the frequency of the summary strategies, the proportion of fixations on AOI3 *Images/graphics supporting content* over the total session, and the surprise emotion are the most important attributes in predicting whether students PASS or FAIL. Students who interact with the ITS with a value higher than 0.03 in the SummAll variable, students who have a proportion of fixations on AOI3 over the total session higher than 0.29, and students who have an emotion of surprise higher than 0.05, are predicted to PASS the course, in other cases they are predicted to FAIL the course.

These results are not surprising considering that Summarizing and Content Coordination of Information Sources are classical strategies that contribute to students taking a strategic approach (Cerezo, Esteban, et al., 2020), and positive emotions such as surprise, enjoyment and happiness are thought to promote motivation, facilitating use of flexible learning strategies, and supporting self-regulation of learning (Pekrun et al., 2011), all of which presumably promote better performance.

## 6   Conclusions

This paper proposes the use of ensembles and attribute selection for improving the prediction of students' performance from multimodal data in an ITS. We collected and preprocessed data from 40 first-year university students from three different sources: learning strategies from MetatutorES logs, emotions from face recording videos, and interaction zones from gaze data, along with marks from performance test about the learning content. We carried out 3 experiments in order to answer two research questions:

- Can attribute selection and classification ensemble algorithms improve the prediction of student final performance from our ITS data? Yes, the use of ensembles and selecting the best attributes approach from numerical data produced the best results in terms of Accuracy and AUC values. The REPTree classification algorithm produced the best results.

- How useful are the models produced and what are the best variables to help teachers understand how to predict students' final performance in the ITS? The white-box models we produced give teachers understandable explanations (IF-THEN rules) of how they arrived at their classifications of student performance. They showed that the attributes that appeared most in these rules were logs denoting use of *Summarizing* strategies and *Coordination of Information Sources* (SummAll and COIStotalFreq) from the ITS logs, paying attention to avatars and to images/graphics supporting text content (AOI2 and AOI6) from gaze data, and surprise from emotions.

The implications of the current study point to Web Intelligent Tutoring Systems and Web-based Adaptive Educational Systems. If data is captured from different data sources, the classifier ensemble methodology proposed in this study could make better, earlier performance predictions than the single data source models that are commonly used at present.

As the next step, we intend to investigate and perform new experiments with the aim of improving our results and in order to overcome some limitations:

- Adding additional different variables/attributes from the multimodal student interaction with the ITS such as think aloud and/or self-report data. In the context of multimodal data, classical self-report methodology remains valuable. Aspects such as achievement emotions experienced by students, students' learning goals and approaches, self-esteem, and epistemological beliefs may help to improve the prediction results. For instance, previous studies have shown that visual metrics (e.g., fixation rate, longest fixations) are significantly influenced by students' goals, so this could be applied to ITS design so that it adapts better to students' learning goals. (Lallé et al., 2017).

  Using EEG (Electroencephalography), ECG (Electrocardiogram), EMG (Electromyography), EDA (Electrodermal Activity), sitting posture, etc. in order to produce more accurate values for predicting students' performance. We would also like to use additional classifier algorithms, particularly deep learning, which could perform significantly better than classic methods.

- Using raw data and other specific data fusion techniques. We used a basic fusion method that uses summary data. However, there are other data fusion theories and methods such as Probability-based methods (PBM) and Evidence reasoning methods (EBM) that we can use with raw data. We could also use semantic (abstract) level features in order to produce intelligent data aggregation.

- We are also aware of the limited generalizability of the results. The next step would be applying the current proposal in other learning systems such as Learning Management Systems (LMSs) or Personal Learning Environments (PLEs). This would allow us to compare results in different learning contexts and with a greater diversity of subjects.

## ACKNOWLEDGMENTS

The authors acknowledge the financial subsidy provided by the Spanish Ministry of Science and Innovation "Deteccion Temprana e Intervencion en Dificultades ee Aprendizaje Especificas desde el Modelo RtI" (PID2019-107201GB-I009), and the Spanish Ministry of Science and Innovation in the project "Improving Data Science User's Experience with Computational Intelligence (INTENSE)" (PID2020-115832GB-I00).

REFERENCES

Adejo, O. W., & Connolly, T. (2018). Predicting student academic performance using multi-model heterogeneous ensemble approach. In *Journal of Applied Research in Higher Education* (Vol. 10, Issue 1, pp. 61–75). https://doi.org/10.1108/JARHE-09-2017-0113

Antonietti, A., Colombo, B., & Di Nuzzo, C. (2015). Metacognition in self-regulated multimedia learning: integrating behavioural, psychophysiological and introspective measures. In *Learning, Media and Technology* (Vol. 40, Issue 2, pp. 187–209). https://doi.org/10.1080/17439884.2014.933112

Arora, R., Sharma, J., Mali, U., Sharma, A., Raina, P., & Professor, A. (2018). Microsoft Cognitive Services. In *International Journal of Engineering Science and Computing*. http://ijesc.org/

Azevedo, R. (2009). Theoretical, conceptual, methodological, and instructional issues in research on metacognition and self-regulated learning: A discussion. *Metacognition and Learning*, *4*(1), 87–95. https://doi.org/10.1007/s11409-009-9035-7

Azevedo, R., Bouchet, F., Harley, J. M., Feyzi-Behnagh, R., Trevors, G., Duffy, M., Taub, M., Pacampara, N., Agnew, L., & Griscom, S. (2011). MetaTutor: An Intelligent Multi-Agent Tutoring System Designed to Detect, Track, Model, and Foster Self-Regulated Learning. *Proceedings of the Fourth Workshop on Self-Regulated Learning in Educational Technologies*, *July*.

Blanchard, E. G., Volfson, B., Hong, Y. J., & Lajoie, S. P. (2009). Affective artificial intelligence in education: From detection to adaptation. *Frontiers in Artificial Intelligence and Applications*, *200*(1), 81–88. https://doi.org/10.3233/978-1-60750-028-5-81

Blikstein, P., & Worsley, M. (2016). Multimodal Learning Analytics and Education Data Mining: using computational technologies to measure complex learning tasks. *Journal of Learning Analytics*, *3*(2), 220–238. https://doi.org/10.18608/jla.2016.32.11

Bogarín, A., Cerezo, R., & Romero, C. (2018). Discovering learning processes using inductive miner: A case study with learning management systems (LMSs). In *Psicothema* (Vol. 30, Issue 3, pp. 322–329). https://doi.org/10.7334/psicothema2018.116

Bondareva, D., Conati, C., Feyzi-Behnagh, R., Harley, J. M., Azevedo, R., & Bouchet, F. (2013). Inferring learning from gaze data during interaction with an environment to support self-regulated learning. *Lecture Notes in Computer Science (Including Subseries Lecture Notes in Artificial Intelligence and Lecture Notes in Bioinformatics)*, *7926 LNAI*, 229–238. https://doi.org/10.1007/978-3-642-39112-5_24

Cerezo, R., Esteban, M., Vallejo, G., Sanchez-Santillan, M., & Nuñez, J. C. (2020). Differential efficacy of an intelligent tutoring system for university students: A case study with learning disabilities. *Sustainability (Switzerland)*, *12*(21), 1–17. https://doi.org/10.3390/su12219184

Cerezo, R., Fernández, E., Gómez, C., Sánchez-Santillán, M., Taub, M., & Azevedo, R. (2020). Multimodal protocol for assessing metacognition and self-regulation in adults with learning difficulties. *Journal of Visualized Experiments*, *2020*(163), 1–24. https://doi.org/10.3791/60331

Cerezo, R., Sánchez-Santillán, M., Paule-Ruiz, M. P., & Núñez, J. C. (2016). Students' LMS interaction patterns and their relationship with achievement: A case study in higher education. In *Computers and Education* (Vol. 96, pp. 42–54). https://doi.org/10.1016/j.compedu.2016.02.006

Chan, M. C. E., Ochoa, X., & Clarke, D. (2020). Multimodal learning analytics in a laboratory classroom. *Intelligent Systems Reference Library*, *158*, 131–156. https://doi.org/10.1007/978-3-030-13743-4_8

Chango, W., Cerezo, R., & Romero, C. (2021). Multi-source and multimodal data fusion for predicting academic performance in blended learning university courses. In *Computers and Electrical Engineering* (Vol. 89). https://doi.org/10.1016/j.compeleceng.2020.106908

Chen, C. M., & Wang, H. P. (2011). Using emotion recognition technology to assess the effects of different multimedia materials on learning emotion and performance. In *Library and Information Science Research* (Vol. 33, Issue 3, pp. 244–255). https://doi.org/10.1016/j.lisr.2010.09.010

Crescenzi-Lanna, L. (2020). Multimodal Learning Analytics research with young children: A systematic review. In *British Journal of Educational Technology* (Vol. 51, Issue 5, pp. 1485–1504). https://doi.org/10.1111/bjet.12959

D'Mello, S., & Kory, J. (2012). Consistent but modest: A meta-analysis on unimodal and multimodal affect detection accuracies from 30 studies. In *ICMI'12 - Proceedings of the ACM International Conference on Multimodal Interaction* (pp. 31–38). https://doi.org/10.1145/2388676.2388686

Greene, J. A., & Azevedo, R. (2010). The measurement of learners' self-regulated cognitive and metacognitive processes while using computer-based learning environments. *Educational Psychologist*, *45*(4), 203–209. https://doi.org/10.1080/00461520.2010.515935

Harley, J. M., Bouchet, F., Hussain, M. S., Azevedo, R., & Calvo, R. (2015). A multi-componential analysis of emotions during complex learning with an intelligent multi-agent system. In *Computers in Human Behavior* (Vol. 48, pp. 615–625). https://doi.org/10.1016/j.chb.2015.02.013

Harley, J. M., Bouchet, F., Papaioannou, N., Carter, C., Trevors, G. J., Feyzi-Behnagh, R., Azevedo, R., & Landis, R. S. (2014). *Assessing Learning with MetaTutor, a Multi-Agent Hypermedia Learning Environment*. https://hal.archives-ouvertes.fr/hal-01217167

Jesus, J., Araujo, D., & Canuto, A. (2017). Fusion Approaches of Feature Selection Algorithms for Classification Problems. In *Proceedings - 2016 5th Brazilian Conference on Intelligent Systems, BRACIS 2016* (pp. 379–384). https://doi.org/10.1109/BRACIS.2016.075

Kuncheva, L. I. (2014). Combining Pattern Classifiers. In *Combining Pattern Classifiers*. https://doi.org/10.1002/9781118914564

Lallé, S., Taub, M., Mudrick, N. V., Conati, C., & Azevedo, R. (2017, June). The impact of student

individual differences and visual attention to pedagogical agents during learning with MetaTutor. In International conference on artificial intelligence in education (pp. 149-161). Springer, Cham.

Lerche, T., & Kiel, E. (2018). Predicting student achievement in learning management systems by log data analysis. In *Computers in Human Behavior* (Vol. 89, pp. 367–372). https://doi.org/10.1016/j.chb.2018.06.015

Li, L. Y., & Tsai, C. C. (2017). Accessing online learning material: Quantitative behavior patterns and their effects on motivation and learning performance. In *Computers and Education* (Vol. 114, pp. 286–297). https://doi.org/10.1016/j.compedu.2017.07.007

Mason, L., Tornatora, M. C., Pluchino, P. (2013). Do fourth graders integrate text and picture in processing and learning from an illustrated science text? Evidence from eye-movement patterns. *Computers & Education. 60*(1), 95-109.

Panadero, E., Klug, J., & Järvelä, S. (2016). Third wave of measurement in the self-regulated learning field: when measurement and intervention come hand in hand. *Scandinavian Journal of Educational Research*, *60*(6), 723–735. https://doi.org/10.1080/00313831.2015.1066436

Pekrun, R., Goetz, T., Frenzel, A. C., Barchfeld, P., & Perry, R. P. (2011). Measuring emotions in students' learning and performance: The Achievement Emotions Questionnaire (AEQ). In *Contemporary Educational Psychology* (Vol. 36, Issue 1, pp. 36–48). https://doi.org/10.1016/j.cedpsych.2010.10.002

Romero, Cristóbal, Romero, J. R., & Ventura, S. (2014). A survey on pre-processing educational data. In *Studies in Computational Intelligence* (Vol. 524, pp. 29–64). Springer Verlag. https://doi.org/10.1007/978-3-319-02738-8_2

Romero, Cristobal, & Ventura, S. (2020). Educational data mining and learning analytics: An updated survey. In *Wiley Interdisciplinary Reviews: Data Mining and Knowledge Discovery* (Vol. 10, Issue 3). https://doi.org/10.1002/widm.1355

Romero, Cristóbal, Ventura, S., & García, E. (2008). Data mining in course management systems: Moodle case study and tutorial. *Computers and Education*, *51*(1), 368–384. https://doi.org/10.1016/j.compedu.2007.05.016

Witten, I. H., Frank, E., & Hall, M. a. (2011). Data Mining:Practical Machine Learning Tools and Techniques second edition. In *Complementary literature None*. https://doi.org/0120884070, 9780120884070

Wolpert, D. H. (2002). The Supervised Learning No-Free-Lunch Theorems. *Soft Computing and Industry*, 25–42. https://doi.org/10.1007/978-1-4471-0123-9_3

Zhong, S. H., Li, Y. H., Liu, Y., Wang, Z. Q., Harvey, P. K., Brewer, T. S., Gandhi, S., Oates, T., Boedihardjo, A. P., Chen, C., Lin, J., Senin, P., Frankenstein, S., Wang, X., Daneshmandi, M.,

Ahmadzadeh, M., Ali, U., Arif, K. S., Qamar, U., … Acm. (2015). Multimodal Assessment of

Teaching Behavior in Immersive Rehearsal Environment - TeachLivE (TM). *Icmi'15: Proceedings of*

*the 2015 Acm International Conference on Multimodal Interaction*, *25*(1), 651–655.

https://doi.org/10.1145/2818346.2823306